\documentclass[prb,superscriptaddress, twocolumn, showpacs,preprintnumbers,amsmath,amssymb]{revtex4}
\usepackage{graphicx}
\usepackage{color}
\usepackage{epstopdf}
\usepackage{bm}
\newcommand{\figurewidth}{\columnwidth}

\begin{document}
\title{Heavy strain conditions in colloidal core-shell quantum dots and their consequences on the vibrational properties from \emph{Ab initio} calculations}

\author{Peng Han}
\affiliation{Max-Planck-Institut f\"ur Festk\"orperforschung, Heisenbergstra{\ss}e 1, D-70569 Stuttgart, Germany}
\affiliation{Department of Physics, Capital Normal University, Beijing 100048, China}
\affiliation{Institut f\"ur Physikalische Chemie, Universit\"at Hamburg, Grindelallee 117, D-20146 Hamburg, Germany}
\author{Gabriel Bester}
\email[E-mail:]{gabriel.bester@uni-hamburg.de}
\affiliation{Institut f\"ur Physikalische Chemie, Universit\"at Hamburg, Grindelallee 117, D-20146 Hamburg, Germany}
\affiliation{Max-Planck-Institut f\"ur Festk\"orperforschung, Heisenbergstra{\ss}e 1, D-70569 Stuttgart, Germany}
\affiliation{The Hamburg Centre for Ultrafast Imaging, Luruper Chaussee 149, D-22761 Hamburg, Germany}

\date{\today}
\begin{abstract}
We preform large-scale \emph{ab initio} density functional theory calculations to study the lattice strain and the vibrational properties of colloidal semiconductor core-shell nanoclusters with up to one thousand atoms (radii up to 15.6~\AA). For all the group IV, III-V and II-VI semiconductors studied, we find that the atom positions of the shell atoms, seem unaffected by the core material. In particular, for group IV core-shell clusters the shell material remains unstrained, while the core adapts to the large lattice mismatch (compressive or tensile strain). For InAs-InP and CdSe-CdS,  both the cores and the shells are compressively strained corresponding to pressures up to 20 GPa. We show that this compression, which contributes a large blue-shift of the vibrational frequencies, is counterbalanced, to some degree, by the undercoordination effect of the near-surface shell, which contributes a red-shift to the vibrational modes. These findings lead to a different interpretation of the frequency shifts of recent Raman experiments, while they confirm the speculated interface nature of the low-frequency shoulder of the high frequency Raman peak.
\end{abstract}
\maketitle

\section{Introduction}
Colloidal semiconductor nanoclusters (NCs) or quantum dots (QDs) have attracted a great deal of attention due to their applications in the fields of optoelectronics, spintronics, photovoltaic or bio-labeling\cite{peng05,clark07,beard07,
klimov07,ithurria07,cirillo14,gaponik10,talapin10,cao00,blokland11,poeselt12} and to their potential to emerge as the key components of the next generation displays \cite{bourzac13}. One important milestone was the synthesis of core-shell quantum dots, which contain at least two semiconductors arrangement in an onion-like geometry \cite{yang99,cao00,baranov03,Chilla08,silva13,dzhagan13,todescato13,cirillo14,jing15}.
In this, the core material is surrounded by a different shell material in order to reduce the influence of the possibly imperfect surface onto the core.
Indeed, semiconductor core-shell NCs with a high photoluminescence quantum efficiency were reported \cite{balet04,peng05,schops06,ithurria07,reiss09,smith09,chen13}.

The theoretical modeling of these type of structures, performed at the level of the effective mass
approximation\cite{jing15}, the $k.p$ method\cite{pistol11}, the tight-binding approach\cite{niquet11,neupane11} or the empirical pseudopotential method\cite{schrier06,luo10} were done assuming perfect, unstrained, and unrelaxed atomic positions. Indeed, colloidal quantum dots in their fluid environment were often considered as unstrained, in direct contrast to their self-assembled (Stransky Krastanov) omologues that are known to  only exist because of the presence of strain. The estimate of the atomic relaxation and ensuing strain is not straight forward in colloidal quantum dots as the use of continuum models\cite{weng07,trallero-Giner10,crut11} or empirical force field \cite{lin14}  would require insights about the surface effects. These are difficult to quantify as the surface relaxes inwards, shortening the bond-length, but these bonds are weakened. Only very recently, {\it ab-initio} large-scale calculations showed that the effect of structural relaxation in colloidal QDs is important \cite{khoo11,han11,han12b,voros13} and a lack of such relaxation leads, for instance, to the appearance of unphysical imaginary vibrational frequencies\cite{han11} and red shifts of vibrational modes\cite{han12b}.

In this work, we perform large-scale \emph{ab initio}  density functional theory (DFT) calculations to study the structural and vibrational properties of core-shell Si(core)-Ge(shell), the inverted Ge-Si, InAs-InP, and CdSe-CdS NCs with radii ranging from 13.5 to 15.6~\AA.
We find that
(i) The shell dictates the atom positions of the entire core-shell NCs. This is especially true for group IV Si-Ge and Ge-Si NCs where we see almost no difference between the bond length distribution of the core-shell NCs and the pure NC made of only shell material. For instance, the Ge core in a core-shell Ge-Si NC is compressed to the bulk Si lattice constant (4\% compression of the bond length).
(ii) Both the core and the shell are compressed in the InAs-InP and in the CdSe-CdS NCs. So the lattice constants of core and shell materials do not undergo the compromise of one having compressive and the other tensile strain, as one may expect.
(iii) The bond-length distribution in the NCs goes from homogeneous (small scattering) throughout the NCs  for our group IV NCs to  inhomogeneous only at the interface for our group III-Vs, to inhomogeneous in the entire shell region in our II-VIs NCs. We speculate that the long-range Coulombic interaction in the more ionic II-VIs is responsible for the large bond distortions.
(iv) The frequency shifts we obtain, compared to the bulk frequencies, for our NCs can all be traced back to two fundamental effects: One being the shift of the modes according to strain (given by the Gr{\"u}neisen parameters), the second being the red-shift created by the undercoordination of the near-surface atoms. Both effects tend to work against each other since the NCs are typically compressively strained and the strain effect leads to a blue-shift (with a positive Gr{\"u}neisen parameter).

\section{Method}
We first construct an un-relaxed NCs by cutting a sphere, centered on an atom for group IVs or a cation for III-Vs and II-VIs with $T_d$ point group symmetry, out of bulk zinc-blende material. Then, the surface atoms with only one nearest neighbor bond are removed and the surface dangling bonds are terminated by hydrogen atoms for group IV atoms and pseudo-hydrogen atoms $H^*$ with a fractional charge of 1/2, 3/4, 5/4, and 3/2 for group VI, V, III, and II atoms, respectively.
These atomic positions are relaxed using DFT in the local density approximation (LDA) and
Trouiller-Martin norm-conserving pseudopotentials with an energy cutoff of 30 Ry for group IV and III-V clusters and 40 Ry
for II-VI clusters\cite{cpmd08}. In order to calculate the vibrational eigenmodes of Cd-VI clusters with up to one thousand atoms, we apply a non-linear core correlation (NLCC) to the Cd atoms instead of including $d$-electrons in the valence. The
calculated transverse optical (TO) and longitudinal optical (LO) frequencies of bulk CdS and CdSe at the $\Gamma$ point using NLCC-LDA and fully including the $d$-electrons in the valence are given in Table.~\ref{table:nlcc}. As shown in this table, the phonon frequencies calculated using NLCC are in close agreement with the full calculations.
\begin{table}
\caption{Comparison of TO and LO frequencies (in cm$^{-1}$) at the $\Gamma$ points for bulk CdS and CdSe between the NLCC calculations and the full calculation including the $d$-electrons in the valence.}
\label{table:nlcc}
\begin{tabular}{lcccc}
\hline
\hline
       & TO$^{NLCC}$ & TO$^{d-state}$ & LO$^{NLCC}$ & LO$^{d-state}$ \\
\hline
CdS    & 247.2   &  246.5  & 296.5  & 299.1 \\
CdSe   & 179.4   &  174.8  & 204.1  & 205.2 \\
\hline
\hline
\end{tabular}
\end{table}

The structures are relaxed until the forces are less than 3$\times$10$^{-6}$ a.u.. The dynamical matrix
elements are then calculated via finite difference and the vibrational eigenmodes and eigenvectors are obtained by
solving the dynamical matrix,\cite{yu10}
\begin{equation}
\label{eq:eigen}
\sum_{J}\frac{1}{\sqrt{M_{I}M_{J}}}\frac{\partial^{2}V({\bm R})}{\partial{\bm R_{I}}\partial{\bm R_{J}}}
{\bm U_{J}}=\omega^{2}{\bm U_{I}}
\end{equation}
where $I$ and $J$ label the atoms, $M$ are the atomic masses, $V({\bm R})$ the potential energy, ${\bm R}$ the atomic positions, ${\bm U}$ the eigenvectors and $\omega$ the vibrational frequencies.
We use \emph{ab initio} DFT implemented in the CPMD code\cite{cpmd08} to optimize the geometry and to calculate the vibrational eigenmodes of the NCs.

In order to analyze the vibrational eigenmodes in terms of the core and the shell (surface) contributions, we calculate the projection coefficients
\begin{equation}
\label{eq:core}
\alpha^{\nu}_{c(s)}=\frac{\sum_{I}^{N_c (N_s)}|\bm X^{\nu}_{I}|^2}{\sum_{I=1}^{N}|\bm X^{\nu}_{I}|^2},
\end{equation}
where, $N_c$, $N_s$ and $N$ are the core, shell (surface), and the total number of atoms; ${\bm X^{\nu}_{I}}$ represents
the three components belonging to atom $I$ and vibrational mode $\nu$ from the 3$N$ component eigenvectors ${\bm U_{I}}$.

To compare the vibrational properties of core-shell NCs with the phonon density of states (DOS) of their corresponding bulk, the phonon DOS of bulk InAs, InP, CdSe, and CdS are calculated via \emph{ab initio} density functional perturbation theory\cite{abinit}.
The computation of bulk materials is performed using the ABINIT code package\cite{abinit} with the same pseudopotentials and the same energy cutoff as their corresponding NCs, and a Monkhorst-Pack $k$-point mesh is taken as 8$\times$8$\times$8. 

To compare our results with Raman spectroscopy measurements, we calculate the Raman intensities using a phenomenological model proposed by Richter \emph{et al.}\cite{richter81}. Based on this model, the Raman intensity of nanostructures is proportional to the projection coefficient of the vibrational modes of the nanostructure onto bulk modes with a relaxation of the wave-vector selection rule\cite{richter81,han12a},
\begin{equation}
\label{eq:raman}
I(\omega)\propto \sum_{n,\nu,\bm{q}}\frac{|C_{n,\bm{q}}^{\nu}|^{2}}{(\omega - \omega^{\nu})^{2}+(\Gamma_0/2)^2},
\end{equation}
where, $C_{n,\bm{q}}^{\nu}$ is the projection coefficient of the NC mode $\nu$ on the bulk mode $n$ with wave vector $\bm{q}$ and $\Gamma_0$ is the natural Lorentzian linewidth (an empirical parameter in our work). The coefficient $C_{n,\bm{q}}^{\nu}$ are summed up to $\Delta q = 1/(2R)$.

\section{Structural properties}
\begin{figure}
\centerline{\includegraphics[width=.8\figurewidth]{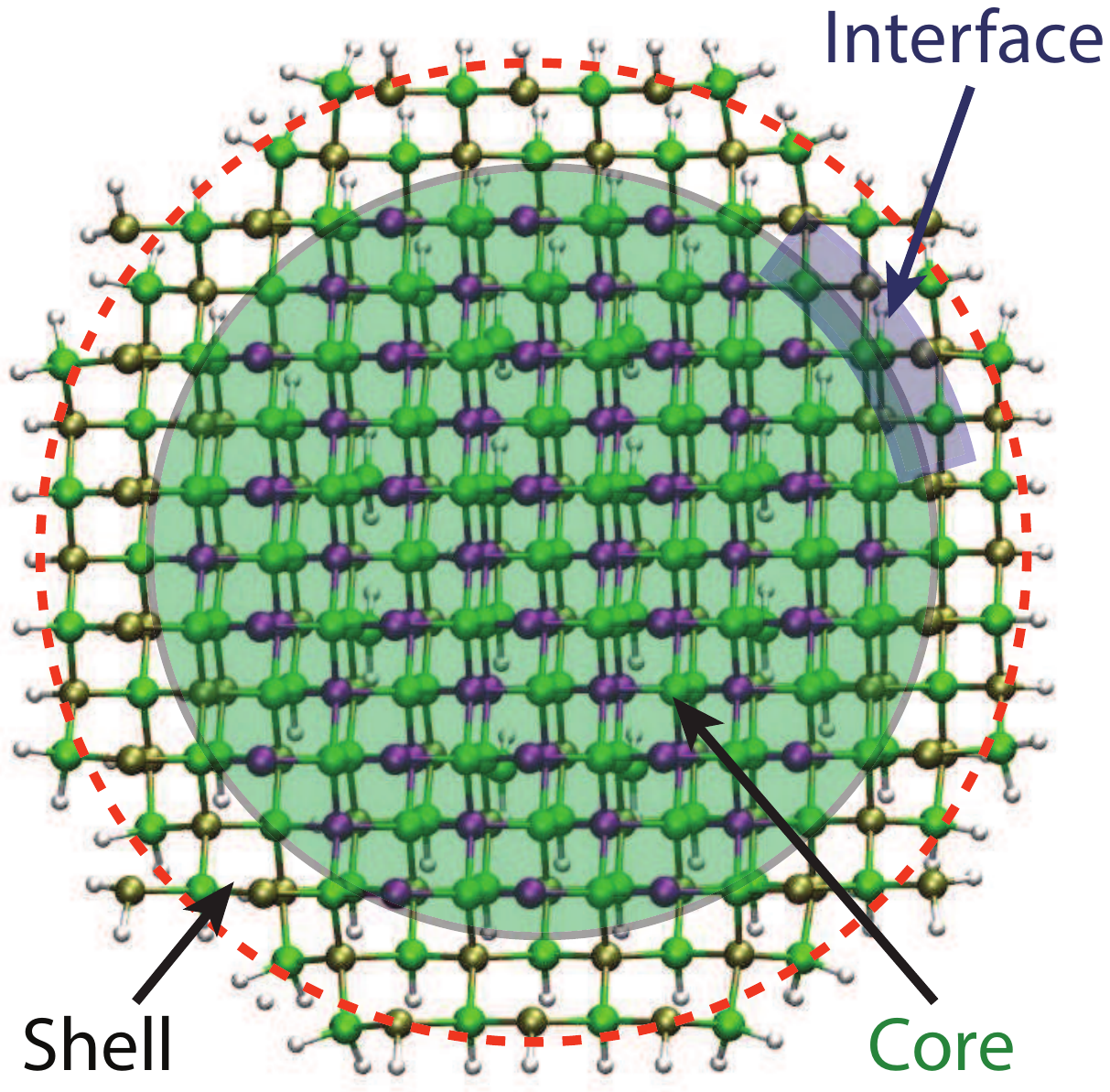}}
\caption{
(Color online) Relaxed atom positions of the In$_{79}$As$_{68}$-In$_{242}$P$_{244}$H$^{*}_{300}$ core-shell nanocrystal. The In, As, P, and H$^*$ atoms are represented as green, purple, tan, and white
spheres, respectively. The definition of core and shell is following the atomic type (As atoms are core-atoms, P-atoms are shell-atoms, in the present case). The interface region is defined graphically.}\label{fig:geom}
\end{figure}

In Fig.~\ref{fig:geom}, we show the relaxed atomic positions of the In$_{141}$As$_{140}$-In$_{228}$P$_{208}$H$^{*}_{300}$ core-shell NC with a radius of 15.6~\AA. In this figure, the core part (InAs) is shown as green and purple spheres while the shell part (InP) is plotted as green and tan spheres. The small white spheres represent the passivants. The core, shell and interface areas are highlighted.

To describe the structural properties of the NCs, we plot the nearest neighbor
distances as a function of their distance from the cluster center. The results are shown for group IV, III-V and II-VI in Figs.~\ref{fig:geom_SiGe},~\ref{fig:geom_InAsP}, and \ref{fig:geom_CdSSe}, respectively.

\subsection{Strain: Group IV NCs}
\begin{figure}
\centerline{\includegraphics[width=\figurewidth]{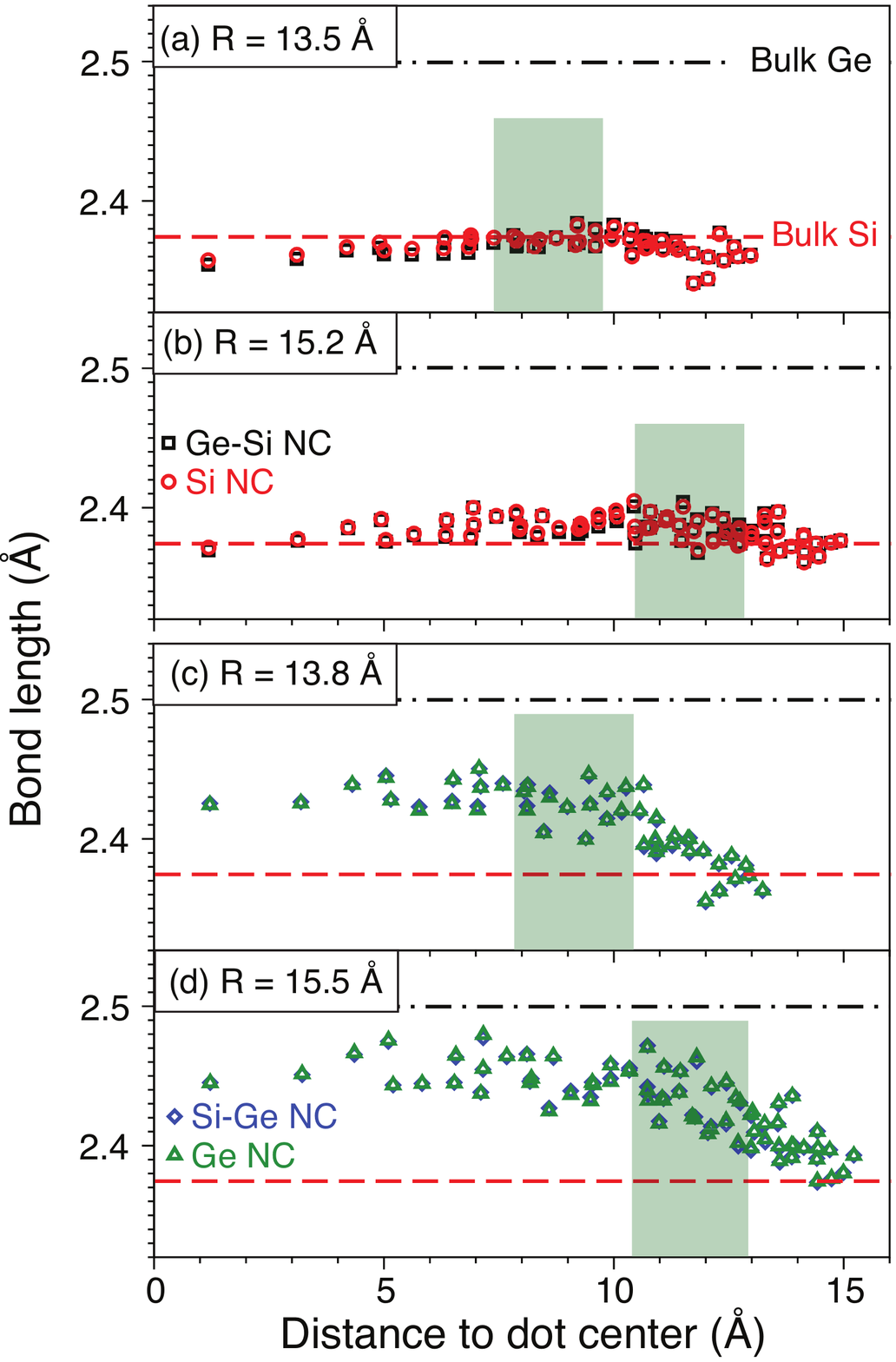}}
\caption{
(Color online)
Bond length distribution as a function of their distance to the dot center. For (a) Si$_{465}$H$_{228}$
(red circles), Ge$_{147}$-Si$_{318}$H$_{228}$ (black squares); (b) Si$_{705}$H$_{300}$
(red circles), Ge$_{281}$-Si$_{424}$H$_{300}$ (black squares); (c) Ge$_{465}$H$_{228}$
(green triangles), Si$_{147}$-Ge$_{318}$H$_{228}$ (blue diamonds); (d) Ge$_{705}$H$_{300}$
(green triangles), Si$_{281}$-Ge$_{424}$H$_{300}$ (blue diamonds). The LDA bond lengths of bulk Si and Ge are given as dashed lines and dotted dashed lines, respectively.
}\label{fig:geom_SiGe}
\end{figure}

For group IV NCs we compare in Fig.~\ref{fig:geom_SiGe} the results for the core-shell structure with the results for a pure NC of similar size made of shell material. The most striking result is that the bond length distribution of both structures is nearly identical. In other words, the core-shell structure is geometrically very similar to a NC made of pure shell material, i.e.,
the shell dictates the structure and the core adapts.

For the {\it Ge-Si core-shell structure}, Fig.~\ref{fig:geom_SiGe}a,b), the softer Ge material (bulk modulus of 77.2 GPa\cite{fine53}) is fully compressed, by around 5\%, to the lattice constant of Si. The bond lengths show a very small deviation from the bulk values, even close to the surface, no noticeable deviation is observed. This is a special feature of Si, compared to the other materials.

For the {\it Si-Ge core-shell structure}, Fig.~\ref{fig:geom_SiGe}c,d), the rather stiff Si core (bulk modulus of 97.6 GPa\cite{hopcroft10}) is expanded by 2-3 \%, in order to almost perfectly match the bond length distribution of a pure Ge NC. The Ge NC, in contrast to the Si NC, has a significant bond-length reduction at the surface (around 5\%) and even for our largest NC (Fig.~\ref{fig:geom_SiGe}d) the bulk bond-length is not recovered at the NC's center. We refer to the bond-length reduction at the surface as the {\it undercoordination effect} \cite{khoo10,han12a}, i.e., the missing atomic partners on the vacuum side lead to an inward surface relaxation, as common in surface physics/chemistry.

We further note that the bond-length variation is rather constant across the interface, marked as light-green area in Fig.~\ref{fig:geom_SiGe}.

\subsection{Strain: Group III-V (InAs-InP) NCs}

\begin{figure}
\centerline{\includegraphics[width=\figurewidth]{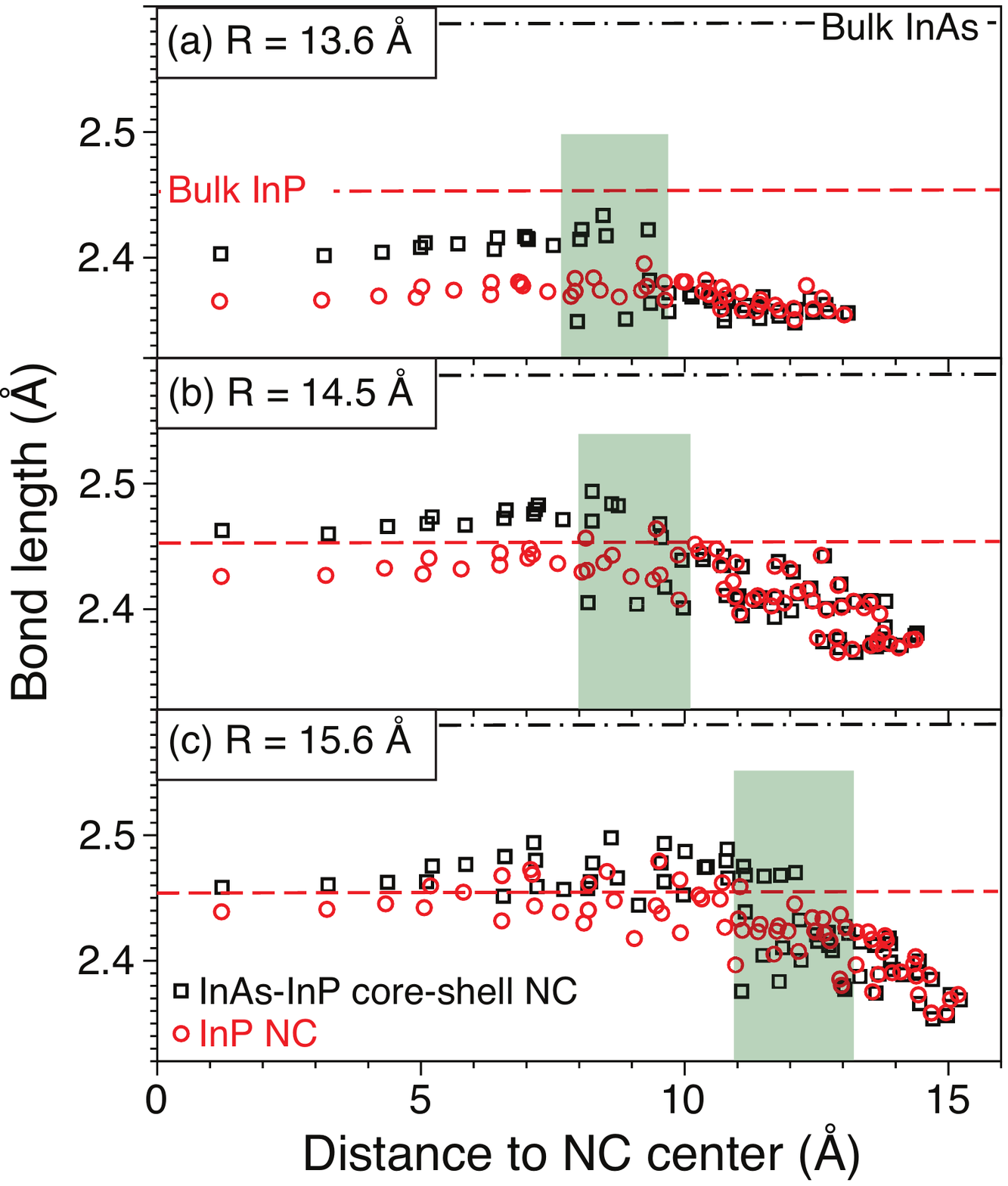}}
\caption{	
(Color online) Bond length distribution as a function of the distance to
the dot center for (a) In$_{79}$As$_{68}$-In$_{146}$P$_{172}$H$^{*}_{228}$ (black square),
In$_{225}$P$_{240}$H$^{*}_{228}$ (red circle);
(b) In$_{79}$As$_{68}$-In$_{242}$P$_{244}$H$^{*}_{300}$ (black square), In$_{321}$P$_{312}$H$^{*}_{300}$ (red circle);
and (c) In$_{141}$As$_{140}$-In$_{228}$P$_{208}$H$^{*}_{300}$ (black square),
In$_{369}$P$_{348}$H$^{*}_{300}$ (red circle). The LDA bond lengths of bulk InP and InAs are given as dashed lines and dotted dashed lines, respectively.}\label{fig:geom_InAsP}
\end{figure}

The corresponding comparison for InAs-InP is given in Fig.~\ref{fig:geom_InAsP}. We now notice a different  bond-length distribution between the InAs-InP core-shell structure and the pure InP NCs.  The difference is, however, still surprisingly small and only apparent for the smaller NCs (Figs.~\ref{fig:geom_InAsP}a,b)). In all cases, the core is heavily compressed, by 5.1\%--7.1\%. For a radius of 16.5 {\AA} (Fig.~\ref{fig:geom_InAsP}c)), the bond-length of the core are already ``converged" to the bond length of the shell. So again, the shell dictates the structure and the core adapts.
The shell in the core-shell InAs-InP NC, or the surface-area in the pure InP NC, have a reduced bond-length due to the undercoordination effect: the surface layers relax inwards.
The bond-length variation is significant in the interface region and hence much larger than in the group IV materials.

\subsection{Strain: Group II-VI (CdSe-CdS) NCs}

\begin{figure}
\centerline{\includegraphics[width=\figurewidth]{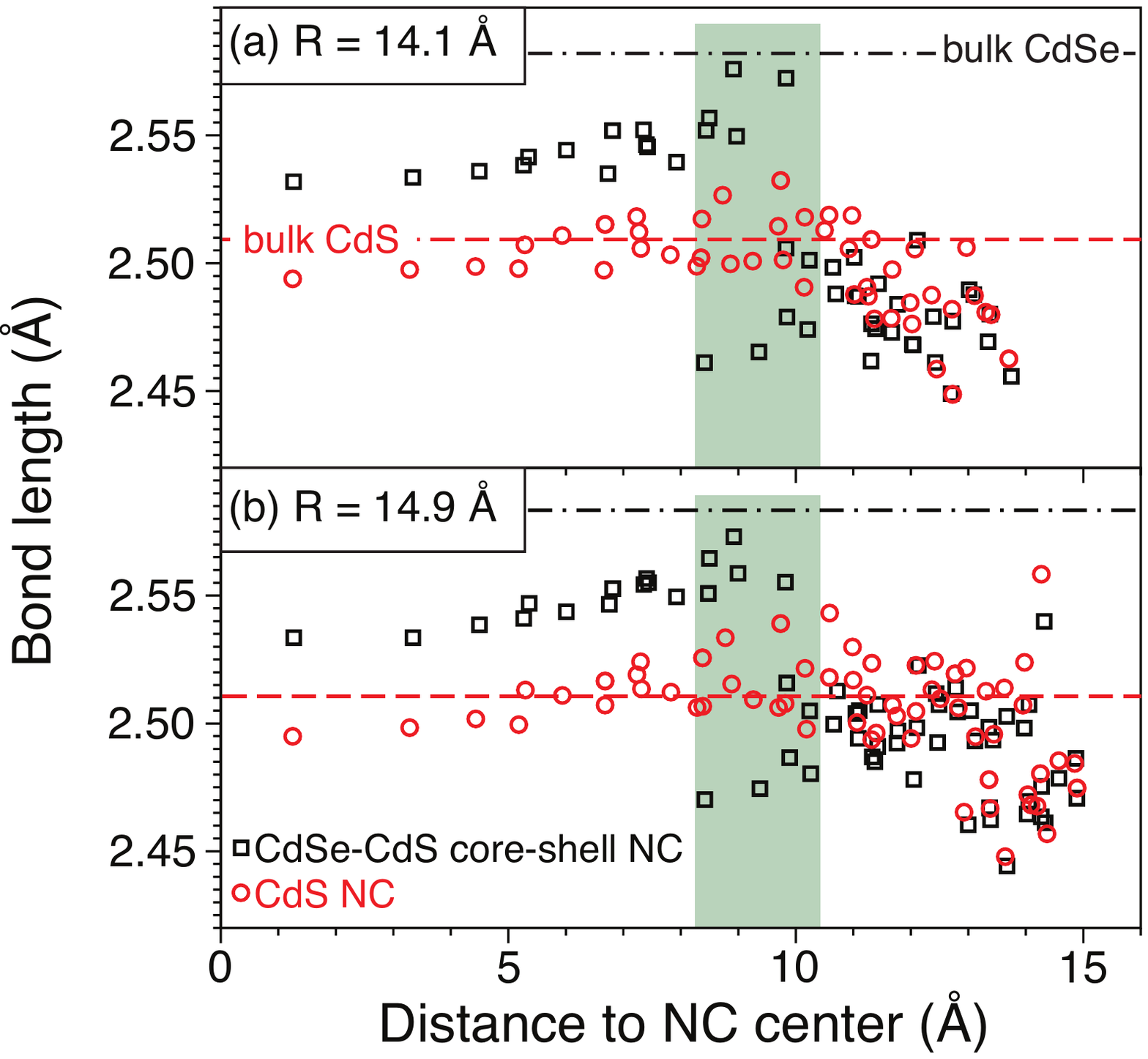}}
\caption{
(Color online) Bond length distribution as a function of the distance to
the NC center for (a) Cd$_{79}$Se$_{68}$-Cd$_{146}$S$_{172}$H$^{*}_{228}$ (black square),
Cd$_{225}$S$_{240}$H$^{*}_{228}$ (red circle),
and (b) Cd$_{79}$Se$_{68}$-Cd$_{242}$S$_{244}$H$^{*}_{300}$ (black square), Cd$_{321}$S$_{312}$H$^{*}_{300}$ (red circle).
The LDA bond lengths of bulk CdS and CdSe are given as dashed lines and dotted dashed lines, respectively.
}\label{fig:geom_CdSSe}
\end{figure}

The results for CdSe-CdS NCs are given in Fig.~\ref{fig:geom_CdSSe}. In comparison to the previous results for our group IV and our group III-V NCs, we can see a clear progression towards a more scattered data: the core-shell structure (black squares) shows a bond-length distribution that significantly differs from the corresponding NC made of only shell material (CdS, red circles). The core CdSe material is still compressed by as much as 2.0\% but has larger bond-length than the core of the corresponding pure CdS NC. In the area of the interface, the bond-length variation is large, going from nearly CdSe bulk bond-length to 1.9\% below the CdS bulk bond length (again, the undercoordination effect).

The comparison of  Fig.~\ref{fig:geom_CdSSe}a) and b) shows as noticeable difference, especially in the surface/shell region, despite the fact that the NCs have only a small difference in radius. This highlights the fact, that the atomistic description leads to a shell-by-shell construction of the structure with increasing radius. A small radius increase can lead to geometrically and chemically rather different structures. The pure CdS NCs have a ratio of Cd to S atoms of 225/240 in the smaller structure and 321/312 in the larger structure. So, Cd poor in the first case and Cd rich in the second. This should be kept in mind when comparing structures with different radii.

\section{Vibrational properties}

The vibrational DOS are shown for the different NCs in Figures.~\ref{fig:SiGe_vib}, \ref{fig:InAsP_vib}, and \ref{fig:CdSeS_vib}.
Two dominant effects lead to the vibrational frequency shifts we observe: (i) {\it Strain induced shifts}. The magnitude of this shift is quantified by the Gr\"uneisen parameter. For the materials considered, these are all positive and between 0.89 and 1.89 (see Table~\ref{tab:gruen}). The Gr\"uneisen parameters for CdS and CdSe TO-modes are results of our DFT calculations, as we did not find experimental results in the literature. This means, for a compression of the structure, the LO and TO frequencies shift to higher frequencies ({\bf blue shift}).
(ii) {\it The undercoordination effect}. The atoms at the surface lacking bonding partners vibrate with a lower frequency. This {\bf red-shift} is strongest for atoms close to the surface but does penetrate a few layers inside the NCs.
\begin{table}[htdp]
\caption{Gr\"uneisen parameters for the optical modes\cite{Landolt01}. The results marked with an asterisk ${}^*$ are from our DFT calculations.}
\begin{center}
\begin{tabular}{l|llllll}
\hline
\hline
      &  Si\cite{Landolt01}    & Ge\cite{Landolt01}    & InP\cite{aoki84}   & InAs\cite{aoki84}   & CdS\cite{wasilik74}   & CdSe\cite{alivisatos88}  \\
\hline
$\gamma^{TO}$  & ~~1.02  & 0.89  & 1.24  & 1.06  &  1.89$^*$   &  1.86$^*$   \\
$\gamma^{LO}$  & ~~1.02  & 0.89  & 1.44  & 1.21  & 1.37 (1.34$^*$) & 1.1 (1.43$^*$)  \\
\hline
\hline
\end{tabular}
\end{center}
\label{tab:gruen}
\end{table}%

\subsection{Vibrations: Group IV NCs}

\begin{figure}
\centerline{\includegraphics[width=\figurewidth]{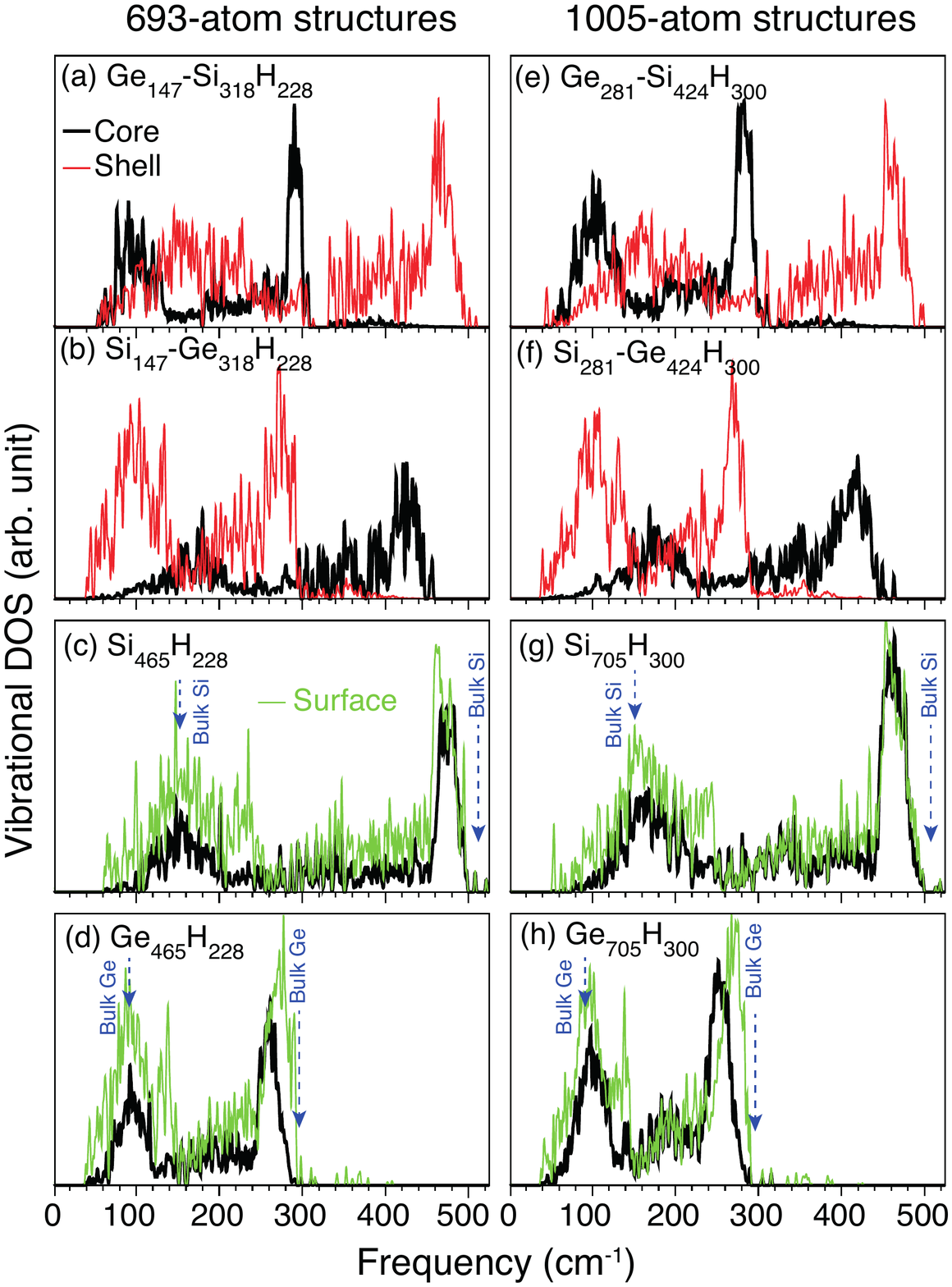}}
\caption{(Color online) Vibrational DOS of
(a) Ge-Si core-shell NC with $R$ = 13.5~\AA, (b) Si-Ge core-shell NC with $R$ = 13.8~\AA, (c) pure Si NC with $R$ = 13.5~\AA, (d) pure Ge NC with $R$ = 13.8~\AA,
(e) Ge-Si core-shell NC with $R$ = 15.2~\AA, (f) Si-Ge core-shell NC with $R$ = 15.5~\AA, (g) pure Si NC with $R$ = 15.2~\AA, (h) pure Ge NC with $R$ = 15.5~\AA. All the vibrational modes were broadened by 0.8~cm$^{-1}$.
The lower frequency blue dashed lines in panels (c), (d), (g), and (h) label the van Hove singularities for the acoustic branches, which corresponds to a maximum in the bulk acoustic phonon DOS.  The higher frequency blue dashed lines show the optical $\Gamma$-point frequencies in bulk (517.0~cm$^{-1}$ for Si and  300.9~cm$^{-1}$ for Ge).
The ``surface" of the pure Si(Ge) NCs are defined to have the same dimension as the ``Shell" in the core-shell NCs. Note that the passivant vibrations are far remote at much higher frequencies, see Ref. \onlinecite{han12b}
for a detailed description.
}\label{fig:SiGe_vib}
\end{figure}

The vibrational DOS is given for  ``smaller" (693 atoms) and ``larger" (1005 atoms) structures in Fig.~\ref{fig:SiGe_vib}. In case of the core-shell NCs (panels (a),(b),(e),(f)), the vibrational eigenmodes are projected onto a core and a shell area defined in Fig.~\ref{fig:geom} and shown as black and red lines respectively in Fig.~\ref{fig:SiGe_vib}.  For the pure NCs (panels (c),(d),(g),(h)) the eigenmodes are projected onto a core and a surface area.

We make few observations:

1) The vibrational DOS of the core-shell structures ((a),(b),(e),(f)) is qualitatively a superposition of the vibrational DOS of the pure Si and Ge NCs ((c),(d),(g),(h)).

2) Compared to the bulk optical frequencies (blue dashed lines at high frequencies in panels (c),(d),(g),(h)), the optical peak is red-shifted. In case of the Si NCs (panels (c),(g)) it is entirely due to the undercoordination effect, as the structure is basically unstrained (see Fig.~\ref{fig:geom_SiGe}) for the bond-length distribution. For the case of the Ge NCs (panels (d),(h)), the undercoordination effect (red-shift) is counterbalanced by the compression (blue-shift) experienced throughout the crystal (see Fig.~\ref{fig:geom_SiGe}). The undercoordination effect is stronger and the optical peak is slightly red shifted.

3) The high frequency optical peak (originating from Si) of the Ge-Si NCs (panels (a), (e)) is significantly blue shifted with respect to the optical peak of the Si-Ge NCs (panels (b), (f)). This shift originates mainly from the fact that the Si core (panels (b)(f)) experiences tensile strain of up to 3\% compared to its nearly unstrained value when it is used as shell material (panels (a) (e)). This is mainly a strain effect.

4)  For Si and Ge, the Gr\"{u}neisen parameters of  the longitudinal acoustic (LA), LO, and TO modes are positive while those of the transverse acoustic (TA) modes are negative.
Due to the mixing of transverse and longitudinal characters in the confined NCs, the
positive and negative Gr\"{u}neisen parameters of the acoustic modes tend to cancel each other out. This leads to the lack of shift in the acoustic modes that we highlighted in Fig. ~\ref{fig:geom_SiGe} (panels (c)(d)(g)(h)) by comparing the NC's results with the
van Hove singularity of the bulk acoustic modes\cite{tubino72} marked with blue dashed lines at 155~cm$^{-1}$ for Si,
and 90~cm$^{-1}$ for Ge.

\subsection{Vibrations: Group III-V (InAs-InP) NCs}

\begin{figure}
\centerline{\includegraphics[width=\figurewidth]{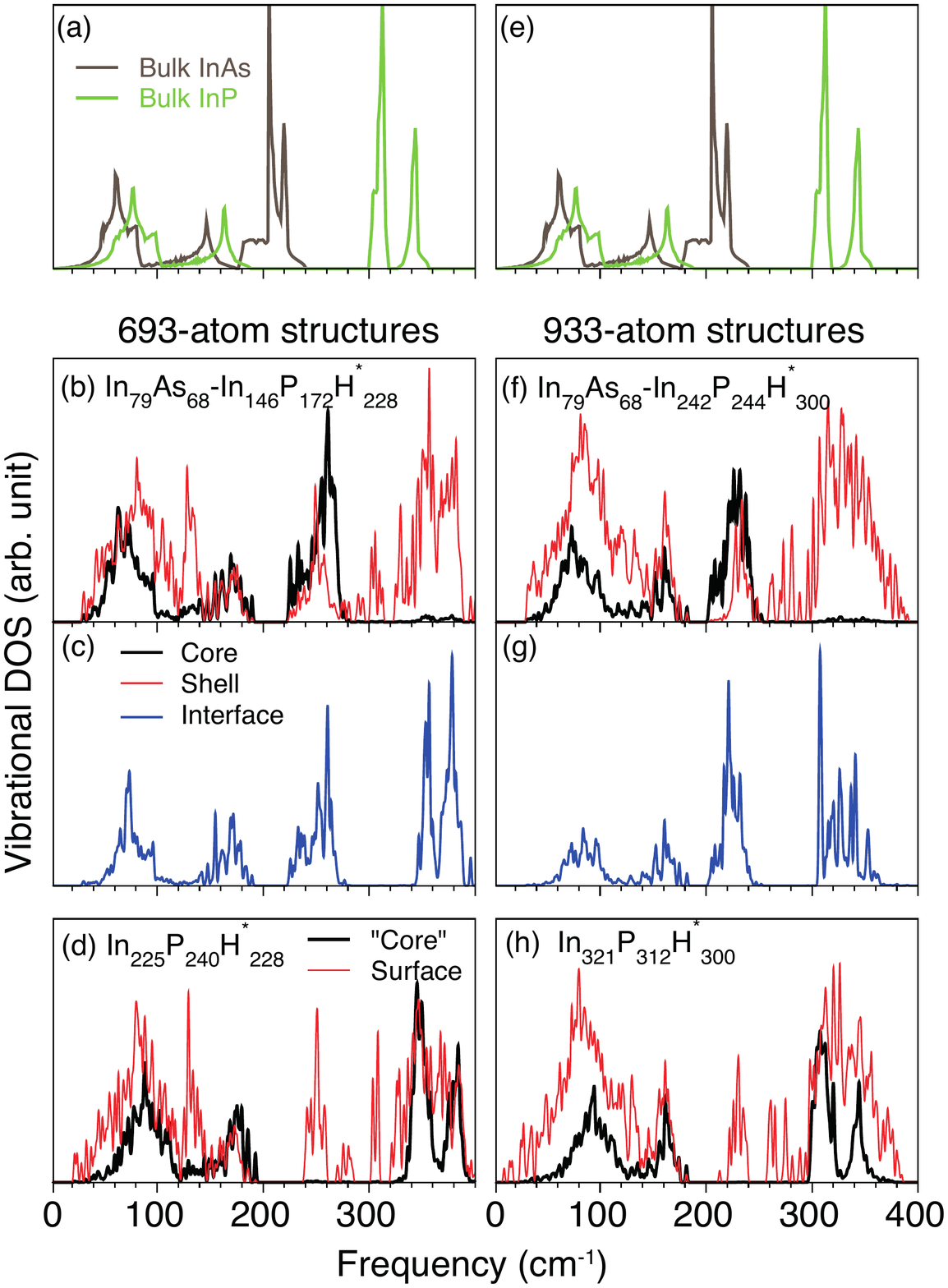}}
\caption{
(Color online) (a) and (e) phonon DOS of bulk InAs and InP (both (a) and (e) panels show the same results). Vibrational DOS of (b) InAs-InP core-shell NC with $R$ = 13.6~\AA, (c) projection on interface region (enlarged 3.25 times), (d) InP NC with $R$ = 13.6~\AA, (f) InAs-InP core-shell NC with $R$ = 14.5~\AA, (g) projection on interface region (enlarged 4 times), (h) InP NC with $R$ = 14.5~\AA. All vibrational modes were broadened by 0.8~cm$^{-1}$. The ``Surface'' of InP NCs is defined to have the same dimension as the ``Shell" in the corresponding core-shell NCs.
}\label{fig:InAsP_vib}
\end{figure}

In a similar fashion as done for group IV, the vibrational DOS is given for InAsP NCs in Fig.~\ref{fig:InAsP_vib}. We note the following:

1) In contrast to the group IV NCs, the optical peaks are strongly blue-shifted compared to the bulk frequencies. This is true for the pure NCs (panels (d) and (h)) and for the core-shell structures (panels (b),(c),(f),(g)). This can be traced back  to the very strong compression given in the core (in the case of the InAs core, more than 6\%) and in the shell (1-2\% in the InP shell), as well as in the pure NC, as shown in Fig~\ref{fig:geom_InAsP}. Such a reduction of 6\% in the bond-length corresponds to extreme pressures of around 20 GPa, for the InAs core. A result that highlights the need to consider strain effects in colloidal QDs.

2) The surface modes that appear in the phonon gap (between acoustic- and optical-type vibrations) between 240-280 cm$^{-1}$ in the InP NCs (panels (d) and (h), see Ref.\onlinecite{han12a} for a detailed description of these new modes) are in the same frequency range as the core optical modes in the core-shell NCs (panels (b) and (f)).

3) Some vibrational modes are confined around the interface (panels (c) and (g)), but lie in the same frequency range as core and shell modes.

\subsection{Vibrations: Group II-VI (CdSe-CdS) NCs}

\begin{figure}
\centerline{\includegraphics[width=\figurewidth]{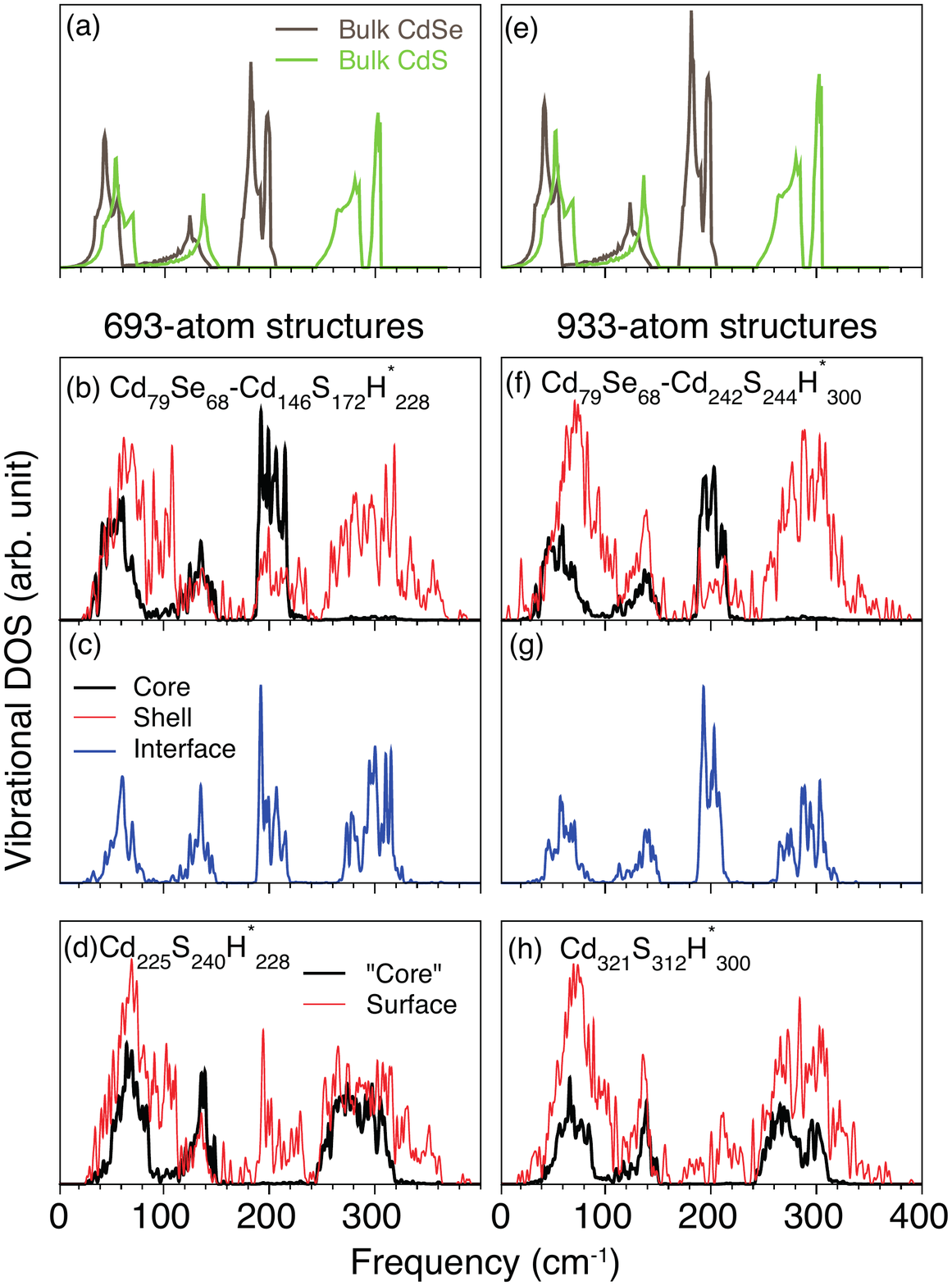}}
\caption{
(Color online) Same as Fig.~\ref{fig:InAsP_vib}, but for CdSe-CdS core-shell NCs with $R$ = 14.1~\AA~[(b)-(d)] and with  $R$ = 14.9~\AA~[(f)-(h)].
}\label{fig:CdSeS_vib}
\end{figure}

In a similar fashion as done for group IV and III-Vs, the vibrational DOS for CdSeS NCs are given in Fig.~\ref{fig:CdSeS_vib}. We note the following:

1) The CdSe core ``optical" modes at around 200 cm$^{-1}$ (panels (b), (f))  are blue-shifted compared to the corresponding bulk frequencies. This is due to a compression of around 2\% of the core (see Fig.~\ref{fig:geom_CdSSe}).

2) The CdS shell ``optical" modes are scattered over a large range of frequencies between 250-370 cm$^{-1}$. This is similar to the case of pure CdS NCs (panels (d) and (h)) and is directly related to the large variation in bond-length approaching the surface of the NCs, as depicted in Fig.~\ref{fig:geom_CdSSe}.

3) Interface modes do exist and are of the persistent-type\cite{Klingshirn12}, i.e., they are in the frequency ranges of the bulk CdSe and CdS modes (with the shifts according to the existing strain). In other words, as is the case for the III-V NCs discussed previously, no additional modes appear in the  intermediate frequency range between the bulk bands.

\section{Comparison with experiment}

\begin{figure}
\centerline{\includegraphics[width=\figurewidth]{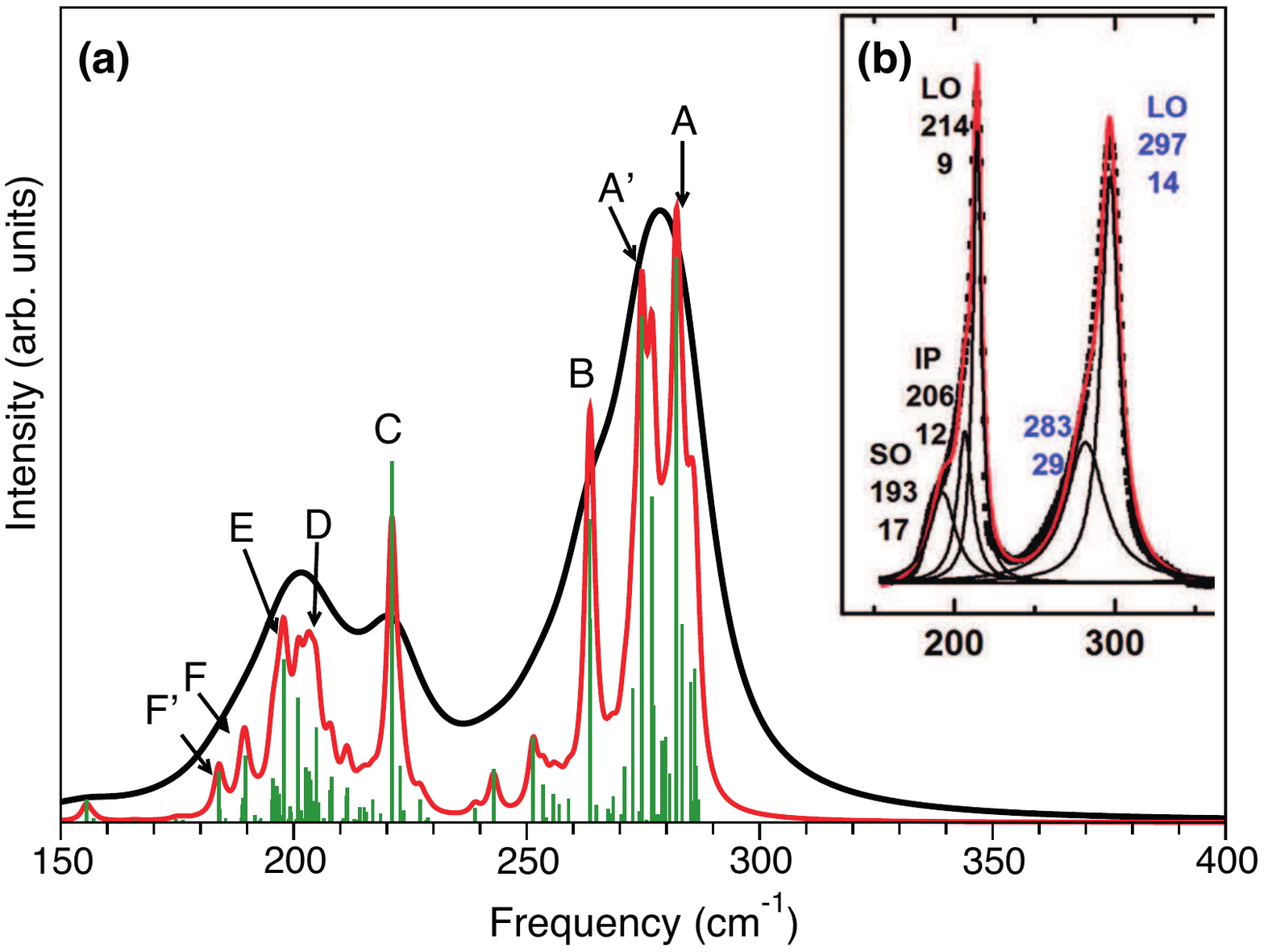}}
\caption{(Color online) (a) Calculated Raman spectrum (see Eq.~(\ref{eq:raman})) for CdSe-CdS core-shell NC with $R$ = 14.9~\AA~. Green lines represent the raw data of the intensities, red and black lines represent the Raman intensities with a broadening of 1.5~cm$^{-1}$ and 8.0~cm$^{-1}$, respectively. (b) Measured Raman spectrum of CdSe-CdS core-shell NC with $R$ around 30.0~{\AA} taken from Ref.~\onlinecite{tschirner12}.
}\label{fig:raman}
\end{figure}

To allow a comparison with experiment we have used the phenomenological model proposed by Richter \emph{et al.}\cite{richter81}, given in Eq.~(\ref{eq:raman}), to calculated the Raman intensities and show the results for CdSe-CdS core-shell Cd$_{79}$Se$_{68}$-Cd$_{242}$Se$_{244}$H$^*_{300}$ NC with $R$ = 14.9~\AA~ in Fig.~\ref{fig:raman}. While the green lines represent the raw data of the intensities, we have also used two different values for the broadening $\Gamma_0$: 1.5~cm$^{-1}$ is shown as red line and a broadening of 8.0~cm$^{-1}$ is shown as black line. The experimental results taken from Ref.~\onlinecite{tschirner12} are given in panel (b) of the figure for comparison.

\begin{figure}
\centerline{\includegraphics[width=\figurewidth]{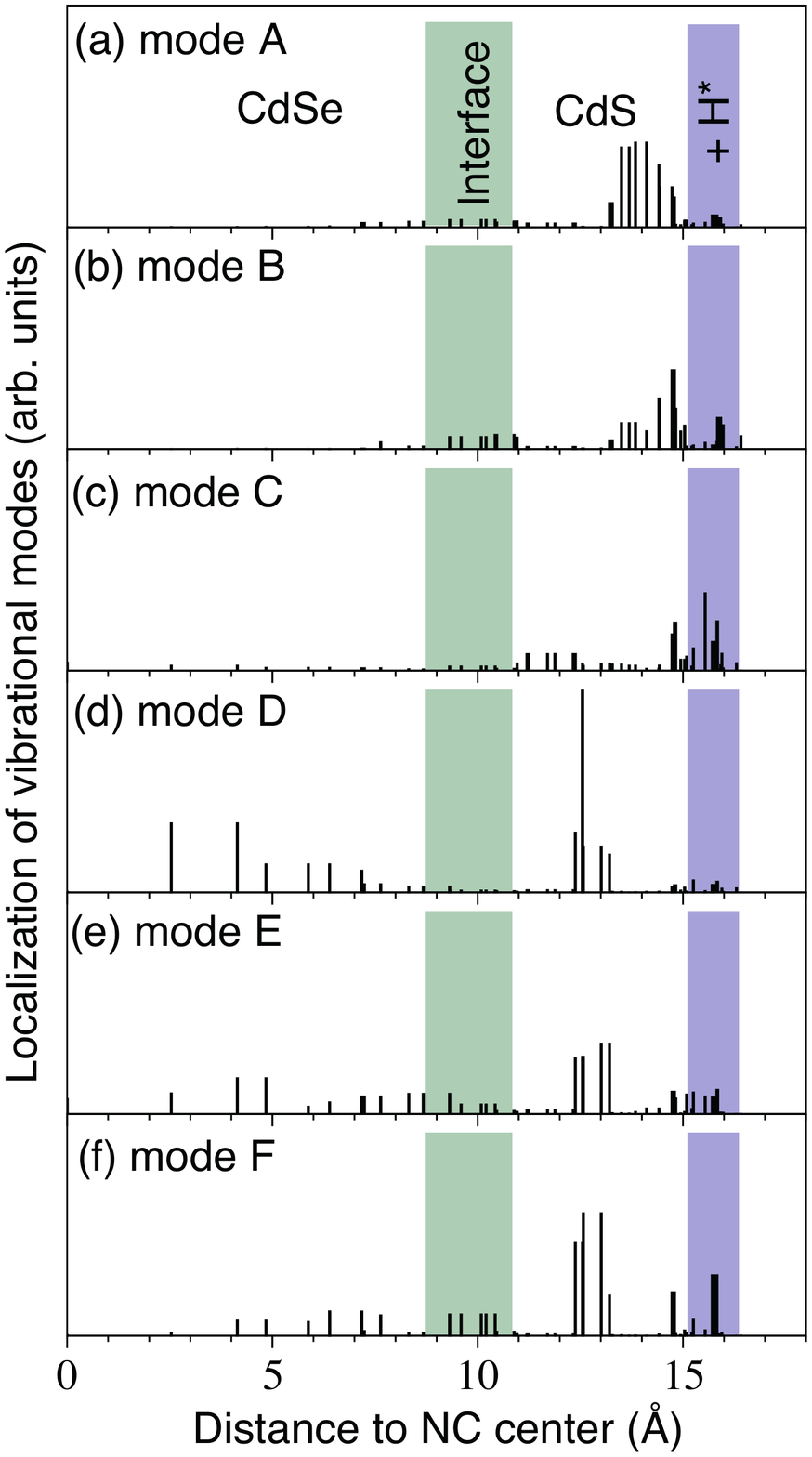}}
\caption{
(Color online) Localization of the vibrational modes of the Cd$_{79}$Se$_{68}$-Cd$_{242}$S$_{244}$H$^*_{300}$ NC with R=14.9~{\AA}, where we have used the magnitude of the eigenmode as indicator (see text). The green highlighted area shows the interface region and the violet area the region where passivant atoms are present defining a surface region.
}\label{fig:vibloc}
\end{figure}

To quantify, as well as possible, the origin of the different vibrations, we have plotted the magnitude of the vibrational eigenvector $|\bm X_I^\nu(R) |$ as a function of its radial position, i.e., distance to the NC center. So for each atom $I$ and eigenmode $\nu$ we obtain one value for $|\bm X_I^\nu(R) |$. The results for all the atoms on one ``shell" around the NC center (with same value of $R$) are averaged and plotted in Fig.~\ref{fig:vibloc}. The vibrations of peak A and A$^{\prime}$, as well as F and F$^{\prime}$ are qualitatively very similar and only the vibrations A and F are plotted.

Two comments are due up front: (1) The size of our NC with a radius of 14.9~{\AA} is significantly smaller than the experimental size with $R$ around 30.0~{\AA}.
(2) {\bf Peak C} (221.4~cm$^{-1}$) is a surface mode of CdS, as can be seen in Fig.~\ref{fig:vibloc} and in Ref.~\onlinecite{supplementary}, which we do not expect to see in the experiment because of the much larger size and corresponding much smaller surface to volume ratio.

Without considering peak C, we observe for the larger broadening (black line in Fig.~\ref{fig:raman}a)) a two peak structure, similar to the experimental result. Each of these peaks is composed of several peaks, as was also deduced from the non-lorentzian line-shapes in the experiment\cite{tschirner12}. We now analyze the peaks subsequently.

{\bf Peaks A} (282.5~cm$^{-1}$) and {\bf A$^{\prime}$} (274.6~cm$^{-1}$) are vibrational modes with optical character of CdS \cite{supplementary}. With a broadening of 8.0~cm$^{-1}$, these two peaks merge into one peak with a frequency of 279~cm$^{-1}$ (Fig.~\ref{fig:raman}), which corresponds to the peak labeled as CdS LO in the experimental work\cite{tschirner12}. Our combined peak is red-shifted compared to the experimental results. From the two effects mainly responsible for frequency shifts, strain and undercoordination, only the latter effect is relevant since the CdS shell is already mainly unstrained in our  NC (see Fig~\ref{fig:geom_CdSSe}). The undercoordination effect becomes weaker with increasing shell size and we expect a blue shift of peaks A and A$^{\prime}$ if we go towards the experimental situation with a much thicker shell, until they reach the bulk value of around 300~cm$^{-1}$ (Fig.~\ref{fig:geom_CdSSe}a)); in good agreement with the experiment. A dependence of the blue shift on the shell thickness was also reported recently \cite{cirillo14}. The experimental peak ``LO 297'' in Ref.~\onlinecite{tschirner12} is therefore a bulk-like, unstrained, CdS peak (see Fig.~\ref{fig:vibloc} to see the localization in the CdS shell) without (large) confinement effect.

{\bf Peak B} (264.3~cm$^{-1}$) is a vibrational mode localized at the interface and in the CdS near-surface shell, with a small but non-vanishing surface component (within the purple area in Fig.~\ref{fig:vibloc}, see Ref.~\onlinecite{supplementary} for a movie of this mode). It corresponds well to the ``low-energy shoulder (LES)" described in the experimental work\cite{tschirner12} and which was observed in NCs with different sizes, shapes, surface environments and materials using Raman and photoluminescence measurements\cite{lin14,roy96,tschirner12,dzhagan11,giugni12,nobile07,hwang99,cherevkov13,venugopal05,onoberov04}. This LES is often described as  a ``surface optical'' phonon mode\cite{lin14,roy96,tschirner12,dzhagan11,giugni12,nobile07,hwang99,cherevkov13,venugopal05,onoberov04} which is in rather good agreement with our identification. The fact, that is some surface character leads to the expectation that it will have some dependence on the type of surface passivation, in good agreement with some experiments \cite{xiong04}.

{\bf Peak D} (203.5~cm$^{-1}$) represents a vibrational mode with optical character of the CdSe core \cite{supplementary} combined with CdS near-interface contributions (see Fig.~\ref{fig:vibloc} (c)) and Ref.~\onlinecite{supplementary}). This mode has a blue shift  (8.5~cm$^{-1}$) compared to the bulk LO mode in CdSe. The experimental blue shift is somewhat larger (214~cm$^{-1}$), which can be traced back to the fact that a structure with a larger shell will experience an even larger compression of the core and hence a larger blue shift.

{\bf Peak E} (197.7~cm$^{-1}$) are modes with contributions from all regions of the NC: core, interface and shell (see Fig.~\ref{fig:vibloc} (d)) and Ref.~\onlinecite{supplementary}). They correspond well to the ``additional intermediate band or IP mode" seen in the experimental work.

{\bf Peak F} (190.0~cm$^{-1}$) and F$^{\prime}$ (184.0~cm$^{-1}$). These modes are CdS shell modes combined with optical surface breathing-type modes (see Fig.~\ref{fig:vibloc} (e)) and Ref.~\onlinecite{supplementary}). They may correspond to the ``SO mode" labeled in the experimental paper.


As a final remark, we note that while the  large majority of theoretical work on core-shell QDs neglect the effect of strain altogether, some used simple continuum models \cite{todescato13,tschirner12} leading to the erroneous conclusion, due to the absence of the surface effect, that the shell structure experiences tensile strain or missed the existence of certain Raman active modes with interface or surface character, such as the LES mode (our mode B). The lack of the latter has forced more intricate interpretation of the experiment in terms of excitonic effects \cite{lin14}, which seem, in view of the present {\it ab initio}  results,  at least not necessary to interpret the experimental results.
A cheap empirical valence force field description would certainly be advantageous, but seem presently out of reach due to the complexity of surface and interface effects. A reasonable valence force field description of the vibrational property of III-V NCs was proposed earlier \cite{han11} for pure NCs. An extension towards core-shell structures and group II-VI or group IV may be worthwhile, although certainly tedious.

\section{Summary}

In summary, we have investigated the structural and vibrational properties of colloidal semiconductor core-shell NCs with up to 1000 atoms via \emph{ab initio} DFT calculations. We find that the geometry of Si-Ge and inverted Ge-Si clusters is determined by the shell part and there is no bond length distortion at the core-shell interface of these group IV NCs. Also for our III-V and II-VI core-shell NCs, the geometry of the shell is similar to that of the corresponding pure NCs made of shell material. Accordingly,  the vibrational DOS of the shell-type vibrations in core-shell NCs  is very similar to the vibrational DOS of pure NCs made of shell material. Hence, the shell experiences no tensile strain but is rather compressed. This fact could be used to improve continuum model descriptions that suggested the opposite situation (tensile shell).

For our III-V and II-VI core-shell NCs we find that the bond-length of the core remains very heavily compressed.
We also find an obvious bond length distortion at the core-shell interface of our III-V and II-VI core-shell NCs. This distortion (i.e., scattering in the bond-length distribution) extends beyond the interface region all the way to the surface in CdSe-CdS core-shell NCs. We link the bond-length distortion to the long range ionic interaction, which is strong in the more ionic II-VI NCs \cite{han12a}. This large scattering of the bond-length distribution leads to a significant broadening of the vibrational bands, which is consequently especially prominent for II-VI NCs.

We also observe a lack of shift in the acoustic modes which we trace back to the mixing of LA and TA modes, with their nearly compensating positive and negative Gr\"{u}neisen parameters \cite{han12a}.
We obtain a blue- (red-) shift in the optical modes of the core parts in Ge-Si (Si-Ge) NCs, which can be traced back to the compressed (expanded) bond lengths.
In addition, we find that the interface modes of III-V and II-VI core-shell NCs are just within the manifold of the core and the shell modes and overlap with the surface modes, which will make them difficult to identify experimentally.

Finally, we compared our results with recent Raman experiments and give a new interpretation of the results based on the lack of tensile strain and the existence of the undercoordination effect. The qualitative picture of our frequency shifts can be understood based on two, often competing, effects of compressive strain (blue-shift of optical modes) and undercoordination (red-shift).  We further show that the often discussed low-energy shoulder on the Raman spectra originate from interface vibrations with small surface character, in agreement with most of the experimental interpretations and in disagreement with earlier theoretical models.

\begin{acknowledgements}
We are grateful to H. Lange and A. Biermann for fruitful discussions. P. H. was supported by the National Natural Science Foundation of China under Grant No. 11404224 and General program of science and technology development project of Beijing Municipal Education Commission under Grant No. KM201510028004. Most of the simulations were preformed on the Cray XC40 Hornet Supercomputer Cluster at the High Performance Computing Center Stuttgart (HLRS).
\end{acknowledgements}

\end{document}